\documentclass[letterpaper,14pt]{article}
\usepackage[]{epsfig}
\usepackage{amsmath}
\usepackage{pstricks}

\begin{document}
        
        \begin{center}
        { Quantum black holes and  the Higgs mechanism at the Planck 
     scale\footnote{This paper is dedicated to the memory of our friend and mentor Antonio Aurilia, who recently passed away}}
        \end{center}

	\begin{center}
{Euro Spallucci\footnote{e-mail address: Euro.Spallucci@ts.infn.it }}
	  \end{center}

\begin{center}
{Dipartimento di Fisica Teorica, Universit\`a di Trieste
and INFN, Sezione di Trieste, Italy}
 \end{center}

\begin{center}
{Anais Smailagic\footnote{e-mail address: anais@ts.infn.it }}
	  \end{center}

\begin{center}
{INFN, Sezione di Trieste, Italy}
 \end{center}
	\vskip 1cm

\begin{center}
\begin{abstract} 
In this paper we present a suitably adjusted Higgs-like mechanism producing black holes  at, and beyond, the Planck energy. 
Planckian objects are difficult to classify either as ``particles'', or as  ``black holes'', since the Compton wavelength
and the Schwarzschild radius are comparable. 
Due to this unavoidable ambiguity,  we consider more appropriate a quantum field theoretical (QFT) approach rather than
General Relativity (GR), which is known to break down as the Planck scale is approached.
\emph{A posteriori}, a connection between the two description can be established for masses large enough with respect 
to the Planck mass, though always describing black holes at the microscopic level. \\
We adopt a  QFT inspired by the Higgs mechanism in the sense that a massive scalar field develops a
non-trivial  vacuum  for $m>\mu_{Pl}$. Exictations around this vacuum are Planckian objects we name ``\emph{black particles}''
to remark the ambiguous identity of these objects, as it has been mentioned above. A black particle  eventually
turns into a ``~quantum black hole~'' when the Schwarzschild radius
becomes \emph{larger} than its Compton wavelength. However, for $m=\mu_{Pl}$ the scalar field is massless at the tree-level,
 but develops
a non-trivial vacuum at one-loop through a Coleman-Weinberg mechanism. In this case, excitations describe Planck mass black particles. 
\end{abstract}
\end{center}
%

\section{Introduction}

Physics at the Planck scale is still an open challenge: the semi-classical approach, where matter is quantized but gravity is not, 
cannot be applied anymore. Even String Theory, which is up to now the only self-consistent way to quantize gravity, does not 
provide fully satisfactory answers to questions about the physical nature of black hole degrees of freedom.\\
 The problems are not only technical, e.g. non-renormalizability of gravity,  but also conceptual in nature. A remarkable
example is the ambiguity in
the very distinction between ``~elementary particles~'' and quantum black holes, whatever
is meant by this term, when the Compton wave length and Schwarzschild radius become comparable.
This energy regime corresponds to a ``~\emph{strong-coupling}~'' phase of the gravitational field where
the \emph{effective coupling} is of order one, i.e. $M^2_{BH}G_N \simeq 1 $. This regime is analogous to $QCD$ \emph{confining} 
phase,  in both cases perturbative techniques fail and the physical nature of dynamical degrees of freedom
is substantially different \footnote{ $QCD$ dynamics can be perturbatively formulated only at high energy, where quarks and gluons 
are weakly coupled. At low energy confinement switches-on and dynamical degrees of freedom are composite hadrons.} . 
By analogy with gluons, one considers ``\emph{gravitons}'' as physical excitations in the weak-coupling  phase. It is tempting
to argue that in the strong-coupling phase the role of hadrons is played by objects similar to black-holes. In a series of recent
papers, black holes has been described in terms of ``graviton condensates'' as a possible realization of the idea of composite
gravitational objects
\cite{Dvali:2010bf,Dvali:2010jz,Dvali:2011nh,Dvali:2011th,Dvali:2012mx,Dvali:2014ila,Dvali:2011aa,Dvali:2012gb,Dvali:2012rt,Dvali:2012en,Dvali:2013vxa,Nicolini:2017hnu,
Casadio:2014vja,Casadio:2015lis}
 Before considering any specific model of ``quantum gravity'', we think  a basic question should be answered: what do 
we physically mean by ``~quantum black hole~'' ? \\
Our answer is that the fundamental description of such an object  should  be in terms of
an ``\emph{uncertain}'' event horizon subject to quantum fluctuations. \\
 This, apparently, ``~obvious~'' consideration has an immediate and substantial consequence: any geometric description of
a quantum black hole, assigning a \emph{definite} position to the event horizon, is inadequate. Quantum
oscillations  completely de-localize the horizon in the vicinity of the  Planck mass.  The horizon ``~freezes~''  at the classical 
position when the mass becomes large enough with respect to the Planck mass, thus making the geometrical description feasible again.
 A quantum
mechanical formulation of the fluctuating horizon has been recently considered in
\cite{Casadio:2013ulk,Casadio:2013uga,Spallucci:2014kua,Spallucci:2015jea,Spallucci:2016qrv,Spallucci:2017aod,Casadio:2016zpl,Casadio:2017nfg}.\\
To avoid possible misunderstanding, we stress that we are referring to Planckian black holes which are not the result 
of a gravitational
collapse of an astrophysical object, but are  produced by a genuine quantum process. \\
In this paper we want to make a step forward from quantum mechanical towards a quantum field theory description.\\
The Higgs mechanism is a cornerstone of the Standard Model of Elementary particles. Its role is instrumental in providing
masses  without spoiling renormalizability of the theory. On the other hand, mass/energy is the source of gravity and it is
intriguing to investigate the possibility of a gravitational role of the Higgs field itself. Many 
papers have studied the Higgs field during the inflationary phase of the early universe \cite{vanTent:2004rc,Kaloper:2008gs}.
Here we speculate on the Higgs field as a source of Planck scale black holes.\\
In Section (\ref{tree}) we introduce an adapted, two-phase, scalar field theory; below the Planck mass the field remains massless,
while above this threshold the field  develops a non-trivial vacuum expectation value and becomes massive. Then, we define
an \emph{effective geometry} induced by this massive object, without resorting to classical Einstein equation. Instead, 
we build the line element starting from the very concept of ``gravitational radius''. For the physical mass of the Higgs
field few times larger than the Planck mass, the geometry is well approximated by the standard Schwartzschild metric.
An interesting result of this approach  is that, even in the classical limit, the horizon entropy keeps memory of the quantum origin
of this object in the form of a logarithmic correction to the area law.\\
In Section(\ref{oneloop}) we extend the tree-level analysis of  Section (\ref{tree}) to include the one-loop level contributions.
It turns out that these quantum corrections
play the dominant role when the classical mass is zero. In this case, the
Coleman-Weinberg mechanism \cite{Coleman:1973jx} provides a non-vanishing vacuum expectation value. The one-loop induced mass can be 
identified with the Planck mass itself. In this case, one finds that the gravitational radius cannot
be shorter than the Planck length.\\
In Section(\ref{end}) we give a brief summary and discussion of the results.

\section{Tree-level Higgs mechanism}
\label{tree}
\begin{center}
\begin{figure}[h!]
\includegraphics[height=8cm]{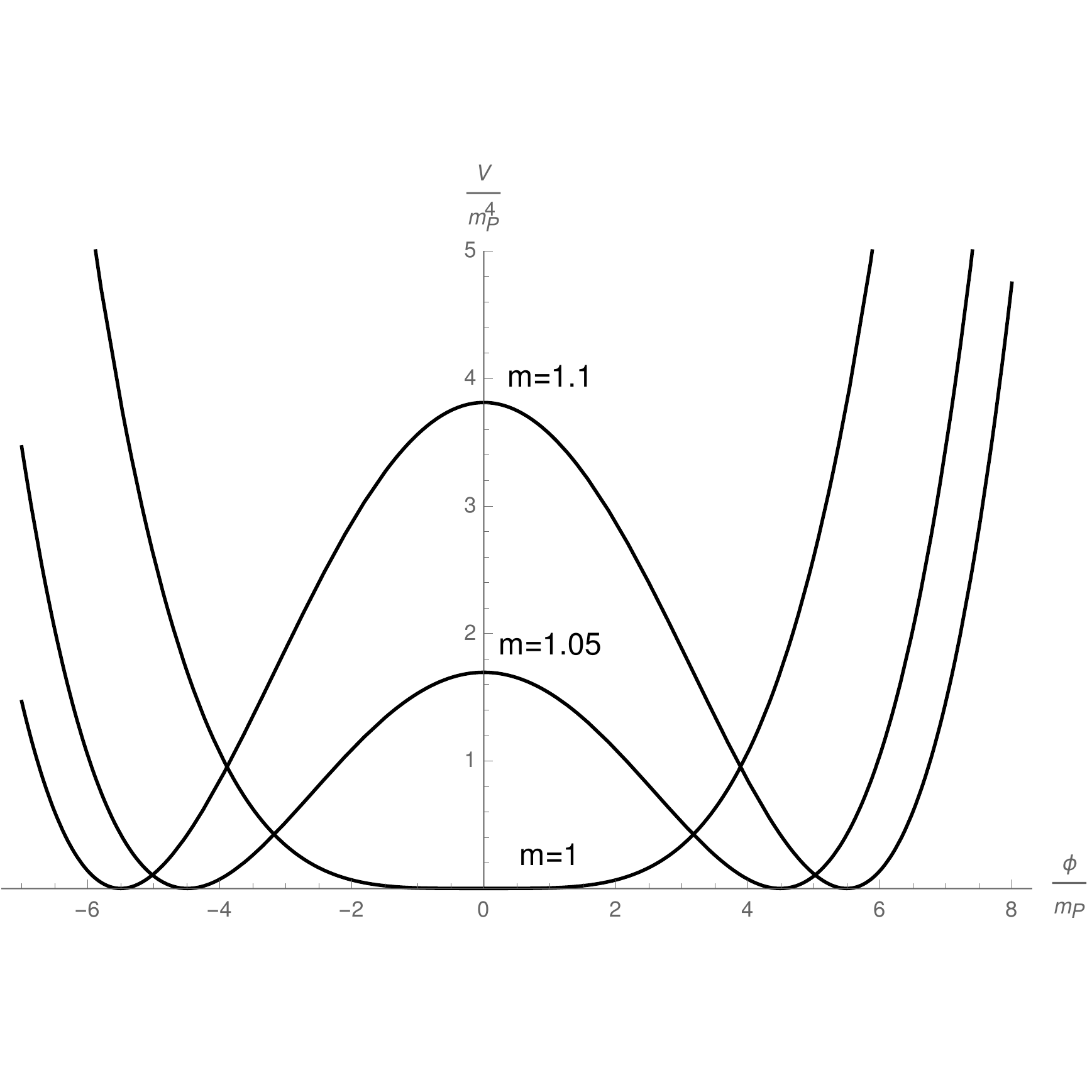}
\caption{Plot of the  classical potential (\ref{hcl}) in Planck units, with $\lambda=0.1$. Curves correspond to
 $m= \mu_{Pl}$; $m= 1.05\mu_{Pl}$; $m=1.1 \mu_{Pl}$ }
\label{clpot}
\end{figure}
\end{center}

One of the heuristic pictures of black hole formation at the Planck scale is through the gravitational collapse of vacuum
energy fluctuations. The problem with this view is that it is described as a \emph{classical} gravitational collapse within a purely
quantum framework. To our knowledge, there is no available  description of the vacuum energy fluctuation  gravitational collapse. \\
An alternative process for micro black hole creation has been formulated in the framework of ``large extra-dimension quantum gravity''
where the unification energy can be lowered down to the $TeV$ scale \cite{ArkaniHamed:1998nn}. In this case,
a ``true'' quantum collapse of a pair of  particles can be represented as an hadronic collision where the impact parameter $b$
is shorter than the effective Schwartzschild radius of the colliding pair, i.e. $b\le 2G_N \sqrt{-s} $, where $s$ is the Mandelstam
invariant mass of the system. Under this condition, the unelastic production channel: hadron $+$ hadron $\longrightarrow $ black hole
can open \cite{Casadio:2008qy,Bleicher:2010qr,Mureika:2011hg,Spallucci:2012xi,Nicolini:2013ega,Casadio:2014pia}. 
Unfortunately,up to now no signal of this process has been observed at LHC.\\  
Here, we present a different possibility for black hole creation through a Higgs mechanism, rather than an hadronic collision. 
Being the way in which elementary
particles get their masses, one can wonder if the same mechanism can also  generate the mass of  
Planckian  black holes. So far, there is no phenomenological evidence of micro black holes with a mass up to $E\simeq 10\, TeV$. 
Thus, the eventual dynamical mechanism producing these structures should activate only above some threshold energy $E^\ast$. 
For the sake of simplicity, we opt for the (very) conservative choice $E^\ast=\mu_{Pl}\simeq 10^{19} GeV$, and consider the simplest
case of a single scalar field with quartic self-interaction and a non-conventional quadratic coupling. The kinetic term has 
the standard form and will be understood in what follows.\\ Let us consider  the potential Fig.(\ref{clpot})

\begin{equation}
 V_{cl}\left(\,\phi\,\right)= -\frac{1}{2}m\sqrt{ m^2 -\mu_{Pl}^2}\,    \phi^2  
+ \frac{\lambda}{4!}\phi^4 +V_0\ ,\quad  0< \lambda << 1
\label{hcl}
\end{equation}

where $\phi$ is a scalar field with canonical mass dimensions; 
$\mu_{Pl}$ is the  Planck mass defined, in natural units $h=1\ , c=1$, as
$2G_N=\mu_{Pl}^{-2}=l_{Pl}^2$, and $V_0$ is a normalization constant in order to guarantee that the vacuum energy
density at the minimum $\phi=\phi_0$  is zero, i.e. $V\left(\,\phi_0\,\right)=0 $. \\
The non-canonical  quadratic coupling is real only for $m \ge \mu_{Pl}$. When the parameter $m$ equals $\mu_{Pl}$
the quadratic term vanishes, the origin $\phi=0$ is the stable minimum and the field excitations around it are massless objects. 
For $m> \mu_{Pl}$ the minimum shifts to a new position $\phi_0\ne 0$

\begin{equation}
 \frac{dV_{cl}}{d\phi}=0 \longrightarrow  \phi\,\left[\,-m\left(\, m^2 -\mu_{Pl}^2\,\right)^{1/2}  
+ \frac{\lambda}{6}\phi^2\,\right]=0
\end{equation}

\begin{eqnarray}
 &&\phi_0=0\longleftrightarrow \quad m\le \mu_{Pl} \label{max} \\
 && \phi_0^2=   \frac{6m }{\lambda}\sqrt{ m^2 -\mu_{Pl}^2} \, 
\longleftrightarrow \quad m >\mu_{Pl}\label{min}
\end{eqnarray}
 In this case, the mass of the field excitations around the
minimum  is given by 

\begin{equation}
 m_{phys}^2 = \left[\, \frac{d^2V_{cl}}{d\phi^2}\,\right]_{\phi=\phi_0}= 2m\sqrt{ m^2 -\mu_{Pl}^2}
\label{hmass}
\end{equation}

The vacuum energy density of the true vacuum is normalized to zero by choosing

\begin{equation}
  V_0  =\frac{3m^2}{2\lambda} \left(\, m^2-\mu_{Pl}^2\,\right)
\label{hvacuum}
\end{equation}

\subsection{ Effective geometry}

Up to this point we proposed an adapted Higgs mechanism with the only novelty that a non-vanishing vacuum expectation value
appears only above some energy scale that we pushed to the Planck mass. We did not mention gravity, space-time curvature,
black holes or General Relativity.  Indeed, the physical mass (\ref{hmass}) can be very small and gravitational effects physically
negligible. In order to have a relevant back-reaction on the surrounding space-time geometry  we expect $m_{phys}\ge\mu_{Pl}$.
Let us consider this point a little more in detail.\\
We start from the basic idea that to each particle of mass $m_{phys}$ one can associate two different length scales:\\
i) the Compton wave length $\lambda_C \equiv 1/m_{phys}$;\\
ii) the ``gravitational radius'' $r_h\equiv 2m_{phys}G_N$.\\
For $\lambda_C> r_h$ we call the object an ``\emph{elementary particle}''; for $\lambda_C< r_h$ we call it a ``\emph{black hole}''.
On the border line $\lambda_C\simeq r_h $, well ... nobody knows!  Some people call these objects 
``~Maximon(s)~''\cite{Markov:1967lha,Markov:1972sc}, others
call them ``precursors''\cite{Calmet:2014gya,Calmet:2015pea}, 
``Planckion(s)''\cite{Aurilia:2002aw,Aurilia:2013psa}, etc.  As an additional contribution to this dictionary of exotic names, we add 
the colloquial term ``\emph{black-particle }''.\\
In spite of its particle-like character, we can still try to associate a ``~metric~'' to a black-particle without referring to
the Einstein equations. It is worth reminding that the idea of gravitational radius predates General Relativity 
\cite{laplace,michell}, and can be used as a starting point to recover an \emph{effective} geometry. This description has
a physical meaning only for distance larger than $l_{Pl}$.\\
To the physical mass (\ref{hmass}) we associate a gravitational radius $r_h$ as:

\begin{equation}
 \frac{r_h^2}{4G_N^2}=\left[\, \frac{d^2V_{cl}}{d\phi^2}\,\right]_{\phi=\phi_0}= 2m\sqrt{ m^2 -\mu_{Pl}^2}\label{effrh}
\end{equation}
Let us notice that $r_h$ will  \emph{eventually } be identified with the radius of the horizon in the limit 
 $m^2\ge 0.5 \left(\ 1 + \sqrt{2}\,\right)\mu_{Pl}^2\equiv m^2_\ast$, where 
 the gravitational length scale is larger than the Compton wavelength. \\
Solving (\ref{effrh}) for $m$ in terms of $r_h$ we find

\begin{equation}
 m=\frac{\mu_{Pl}}{\sqrt{2}}\left(\, 1 + \sqrt{1 + \frac{r_h^4}{l_{Pl}^4}}\,\right)^{1/2}\label{mrh}
\end{equation}

From (\ref{mrh}) we construct the effective metric as

\begin{eqnarray}
 && ds^2=-f(r)dt^2 +f(r)^{-1}dr^2 + r^2 d\Omega^2\ ,\label{bp1}\\
 &&  f(r)= 1 - \frac{\sqrt{2}m}{\mu_{Pl}}\left(\, 1 + \sqrt{1 + \frac{r^4}{l_{Pl}^4}}\,\right)^{-1/2}\ ,\quad m> m_\ast \ ,
r>> l_{Pl}
\label{bp2}
\end{eqnarray}

At large distance, $r>> l_{Pl}$, the metric coincides with the Scwarzschild one:

\begin{equation}
 f(r)\approx  1 - \frac{\sqrt{2}m}{\mu_{Pl}}\frac{l_{Pl}}{r}= 1 -2G_N \frac{\sqrt{2}m}{r}\label{Schw}
\end{equation}


It is customary to describe a black hole in thermodynamical terms. The first quantity to be introduced is 
the Hawking temperature. In our case we find

\begin{equation}
 T_H = \frac{1}{4\pi l_{Pl}^4}\frac{1}{ 1 + \sqrt{1 + r_h^4/l_{Pl}^4 } }\frac{r_h^3}{\sqrt{1 + r_h^4/l_{Pl}^4 }}
\label{tcl}
\end{equation}

$T_H$ approaches the standard form of the Hawking temperature for $m $ large with respect $\mu_{Pl}$. In this case 
$r_h >>l_{Pl} $ and

\begin{equation}
 T_H \longrightarrow \frac{1}{4\pi r_h}
\end{equation}

as it is expected. Furthermore, as $m$ decreases, $T_H$ reaches a maximum value $T_H \approx 0.03 T_{Pl}$ at $r_H=1.59\,l_{Pl}$.
For  $ r_h \to l_{Pl}$, the temperature decreases to $T_H \approx 0.02 T_{Pl}$. \\
This is how far we can push $T_H$ to have physically meaningful results. If one \emph{formally} considers the limit $r_h\to 0$
($m\to \mu_{Pl}$) one gets

\begin{equation}
 T_H \approx \frac{r_h^3}{8\pi l_{Pl}^4} \longrightarrow 0
\end{equation}

This result has to be taken with care: physically it means that we are approaching the region where particles and black holes
are indistinguishable and the thermodynamical description loses its meaning.\\
Another interesting thermodynamical quantity is the entropy, given by the First Law as:

\begin{equation}
 d S = \frac{2\pi}{l_{Pl}}\left(\, 1+ \sqrt{ 1 + r_h^4/l_{Pl}^4}\,\right)^{1/2} dr_h \label{1law}
\end{equation}

Integration of (\ref{1law}) is non-trivial, but can be carried out in the limit $r_h > l_{Pl}$, where we find

\begin{equation}
 S \simeq \pi \left[\, r_h \sqrt{ r^2_h + l_{Pl}^2} 
      + l_{Pl}^2\ln\left(\, \frac{r_h +  \sqrt{ r^2_h + l_{Pl}^2}}{\,l_{Pl}} \,\right)\,\right]\label{lns}
\end{equation}

The result {\ref{lns}} shows the appearance of a logarithmic correction with respect the area contribution. Furthermore,
even in the ``~classical limit~'' $m>> \mu_{Pl}$, this correction survives giving:

\begin{equation}
 S\longrightarrow \pi r_h^2 + \frac{\pi}{2} l_{Pl}^2\ln\left(\, \frac{4r_h^2}{l_{Pl}^2}\,\right)
\end{equation}
The first term is the celebrated Area Law, $S=A_H/ 4$, while  the logarithmic correction can be traced to the quantum nature
of black particles. \\
So far, we have considered only the tree-level Higgs potential (\ref{hcl}). It is possible to further include
one-loop corrections. These corrections are physically negligible in the regime $m> \mu_{Pl}$, but 
become dominant for $m= \mu_{Pl}$ due to the absence of the quadratic term in the tree-level potential.\\

\section{One-loop effects}
\label{oneloop}
\begin{center}
\begin{figure}[h!]
\includegraphics[height=8cm]{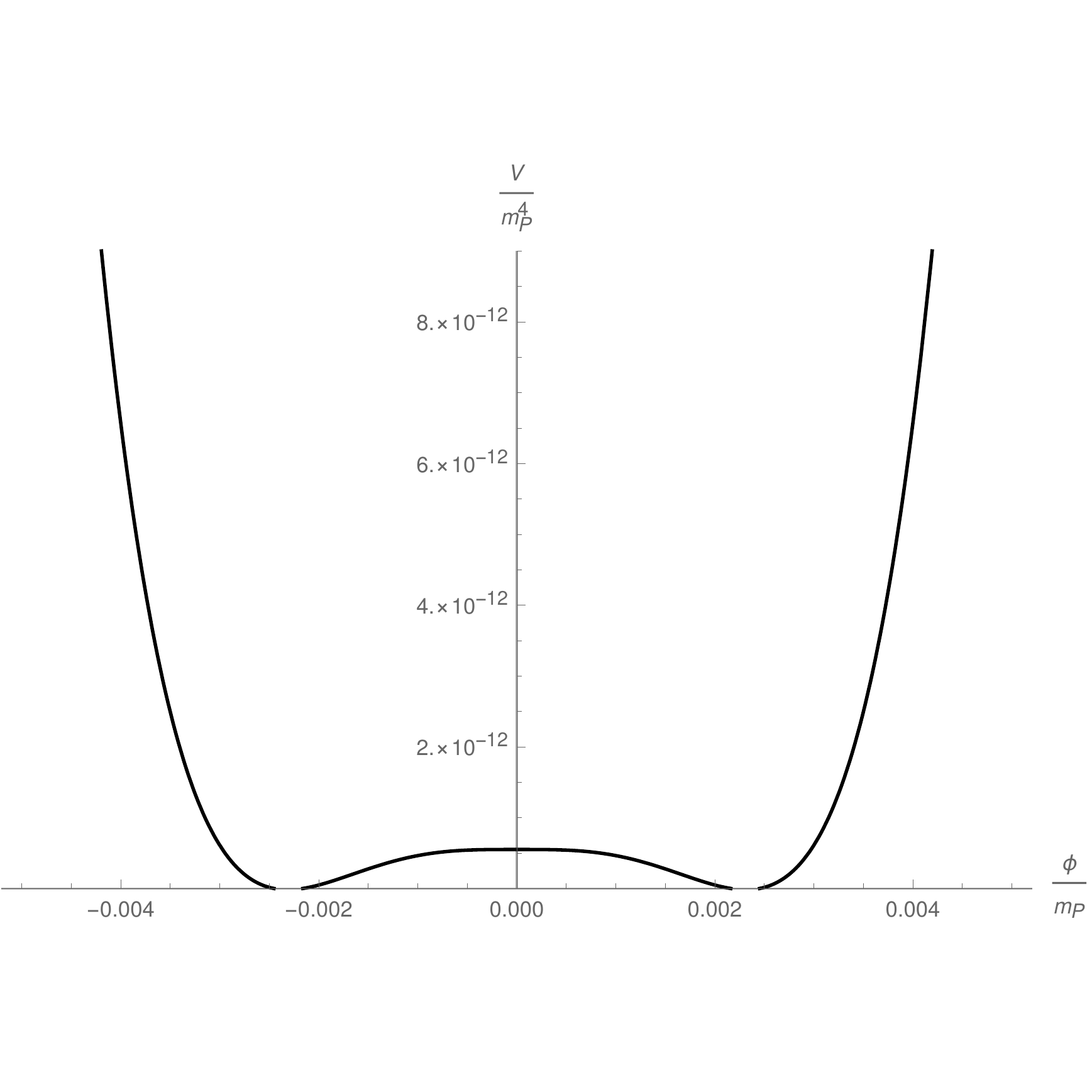}
\caption{Plot of the  Coleman-Weinberg potential  in Planck units, with $\lambda=0.1$
and $m=\mu_{Pl}$.}
\label{cwpot}
\end{figure}
\end{center}

The Higgs mechanism works at the tree-level and quantum corrections do not alter in a significant way the classical results.
However, starting from a  quartic potential, no ``wrong sign'' quadratic term, a non-vanishing vacuum expectation value and
mass can be dynamically generated through the Coleman-Weinberg effect \cite{Coleman:1973jx}. In our case, this would correspond
to start with $m=\mu_{Pl}$. In order to include this case,  
we shall consider, in tis section, one-loop corrections to the tree-level potential. This is a standard calculation which
can be found in quantum field theory textbooks and will not be repeated here.
The general form  of the one-loop effective potential is given by

\begin{equation}
 V_1\left(\,\phi\,\right)= V_{cl}+ \frac{V_{cl}^{\prime\prime\,2}}{64\pi^2}\left(\, \ln \frac{V_{cl}^{\prime\prime}}{\mu^2}
-\frac{1}{2}\,\right)
\end{equation}

where, $\mu$ is the renormalization scale. The order $\hbar$ corrected potential reads in our case Fig.(\ref{cwpot}):

\begin{eqnarray}
 V_1\left(\,\phi\,\right)&&=V_0-\frac{1}{2}m\sqrt{ m^2 -\mu_{Pl}^2}\phi^2 +\frac{\lambda}{4!}\phi^4\nonumber\\
 &&+\frac{1}{64\pi^2}
\left(\, m\sqrt{ m^2 -\mu_{Pl}^2} + \frac{\lambda}{2}\phi^2\,\right)^2\left[\,
\ln\left(\, -\frac{m}{\mu^2}\sqrt{ m^2 -\mu_{Pl}^2} + \frac{\lambda\phi^2}{2\mu^2}\,\right)-\frac{1}{2}\,\right]\nonumber\\
&&\label{v0}
\end{eqnarray}

The quartic coupling constant  is generally assumed to be small, i.e. $\lambda << 1$. Thus, 
 order $\lambda^2$  one-loop corrections are smaller than the tree-level term. In this regime, the previous analysis is essentially 
not significantly modified by quantum effects. This is true for $m> \mu_{Pl}$, but for $m=\mu_{Pl}$ the minimum of the quantum
corrected potential is no more $\phi=0$.\\
Let us look for the extremal points of (\ref{v1}) by equating to zero the first derivative of $V_1(\phi)$:

\begin{eqnarray}
 \left[\,\frac{dV_1}{d\phi}\,\right]_{\phi=\phi_0}=&&-m\sqrt{ m^2 -\mu_{Pl}^2}\phi_0 +\frac{\lambda}{6}\phi_0^3\nonumber\\
&&+\frac{\lambda\phi_0}{32\pi^2}\left(\,-m\sqrt{ m^2-\mu^2_{Pl}}  + \frac{\lambda}{2}\phi^2_0\,\right)
\ln\left(\, -\frac{m}{\mu^2}\sqrt{ m^2 -\mu_{Pl}^2} + \frac{\lambda\phi_0^2}{2\mu^2}\,\right) \nonumber\\
&&=0\label{v1}
\end{eqnarray}

The physical mass is defined as the second derivative of $V_1$ at the eventual minimum:
\begin{equation}
 \left[\,\frac{d^2V_1}{d\phi^2}\,\right]_{\phi=\phi_0} =-m\sqrt{ m^2 -\mu_{Pl}^2} +\frac{\lambda}{2}\phi_0^2
+\frac{\lambda M^2}{32\pi^2}\ln\frac{M^2}{\mu^2}+ \frac{\lambda^2 \phi_0^2}{32\pi^2}\ln\frac{M^2}{\mu^2} + 
\frac{\lambda^2 \phi_0^2}{32\pi^2}\label{v2}
\end{equation}

where $M^2$ is a shorthand for the second derivative of the tree-level potential in $\phi_0$, i.e.
 $M^2\equiv -m \sqrt{ m^2 -\mu_{Pl}^2}  + \lambda\phi^2_0/2$. \\
We can use equation (\ref{v1}) to get rid of the arbitrary mass scale $\mu$ as follows:

\begin{equation}
\frac{\lambda}{32\pi^2}
\ln\left(\, -\frac{m}{\mu^2}\sqrt{ m^2 -\mu_{Pl}^2} + \frac{\lambda\phi_0^2}{2\mu^2}\,\right)
=\frac{ -m \sqrt{ m^2 -\mu_{Pl}^2}  -\frac{\lambda}{6}\phi_0^2}{M^2(\phi_0)} 
\end{equation}

Now we can replace the logarithmic term in (\ref{v2}):

\begin{eqnarray}
\left[\,\frac{d^2V_1}{d\phi^2}\,\right]_{\phi=\phi_0} &&  = -m \sqrt{ m^2 -\mu_{Pl}^2}+\frac{\lambda}{2}\phi_0^2 
+\frac{\lambda^2}{32\pi^2}\phi_0^2
     -\frac{\lambda}{6}\phi_0^2 + m \sqrt{ m^2 -\mu_{Pl}^2}\nonumber\\
&&+\frac{	\lambda\phi_0^2}{M^2(\phi_0)}\left(\,m \sqrt{ m^2 -\mu_{Pl}^2}-\frac{\lambda}{6}\phi_0^2 \,\right)\nonumber\\
 &&=\frac{\lambda}{3}\phi_0^2++\frac{\lambda^2}{32\pi^2}\phi_0^2
+\frac{	\lambda\phi_0^2}{M^2(\phi_0)}\left(\,m \sqrt{ m^2 -\mu_{Pl}^2}-\frac{\lambda}{6}\phi_0^2 \,\right)\label{d2}
\end{eqnarray}

By defining the one-loop physical mass as:
\begin{equation}
 \left[\,\frac{d^2V_1}{d\phi^2}\,\right]_{\phi=\phi_0} =
m^2_{phys}\equiv  2m \sqrt{ m^2 -\mu_{Pl}^2} +\mu_{Pl}^2 \label{mass1}
\end{equation}

we see that:

\begin{itemize}
 \item for $ m^2 -\mu_{Pl}^2 >> \mu^2_{Pl} $  (\ref{mass1}) reproduces equation (\ref{hmass});
\item for $ m = \mu_{Pl} $ it  gives raise to Planck mass black-particle.
\end{itemize}

By inserting (\ref{mass1}) into (\ref{d2}) we obtain an algebraic equation for the minimum :

\begin{eqnarray}
 &&\frac{\lambda^3}{64\pi^2} \phi_0^4 +\frac{2\lambda}{3} m \phi_0^2 \sqrt{ m^2 -\mu_{Pl}^2} 
-\frac{\lambda}{2}m^2_{phys}\phi_0^2\nonumber\\
&&-\frac{\lambda^2}{32\pi^2} m \phi_0^2 \sqrt{ m^2 -\mu_{Pl}^2} 
+m^2_{phys} m \sqrt{ m^2 -\mu_{Pl}^2} =0 \label{minima}
\end{eqnarray}

We stress that the quartic term, originating from the one-loop contribution, is relevant \emph{only} 
very close to $\mu_{Pl}$. 
The limiting case $m=\mu_{Pl}$ gives 
\begin{equation}
  \frac{\lambda^3}{64\pi^2} \phi_0^4  -\frac{\lambda}{2}\mu_{Pl}^2\phi_0^2 =0\Rightarrow 
\phi_0^2 =\frac{32\pi^2}{\lambda^2}\mu_{Pl}^2
\end{equation}
A non-zero vacuum expectation value $\phi_0$ is generated by the Coleman-Weinberg mechanism.\\
As $m$ increases the quartic term   in (\ref{minima}) becomes negligible and
the equation reduces to a quadratic one

\begin{equation}
 \frac{2\lambda}{3}m\sqrt{ m^2 -\mu_{Pl}^2}\phi_0^2-\lambda m\sqrt{ m^2 -\mu_{Pl}^2} \phi_0^2 
+2m^2\left(\, m^2-\mu^2_{Pl}\,\right) =0
\end{equation}

The final result is the tree-level minimum (\ref{min})

\begin{equation}
\phi_0^2 =\frac{6m^2}{\lambda}\sqrt{ m^2 -\mu_{Pl}^2}
\end{equation}

As $m$ approaches $\mu_{Pl}$ form above, i.e.   $m\to\mu_{Pl}$, one has to keep also the order $\lambda^0$ correction.
 The result is

\begin{equation}
 \phi_0^2\simeq \frac{6}{\lambda}m\sqrt{ m^2 -\mu_{Pl}^2}\left(\, 1 -\frac{3\lambda}{32\pi^2}\,\right)
\end{equation}

At this point, it is tempting to see as the one-loop correction modify the effective metric (\ref{bp1}), (\ref{bp2}).
The gravitational length equation (\ref{effrh}) is now

\begin{equation}
 \frac{r_h^2}{4G_N^2}= 2m\sqrt{ m^2 -\mu_{Pl}^2} + \mu_{Pl}^2 \label{effrh1}
\end{equation}

\begin{center}
\begin{figure}[h!]
\includegraphics[height=8cm]{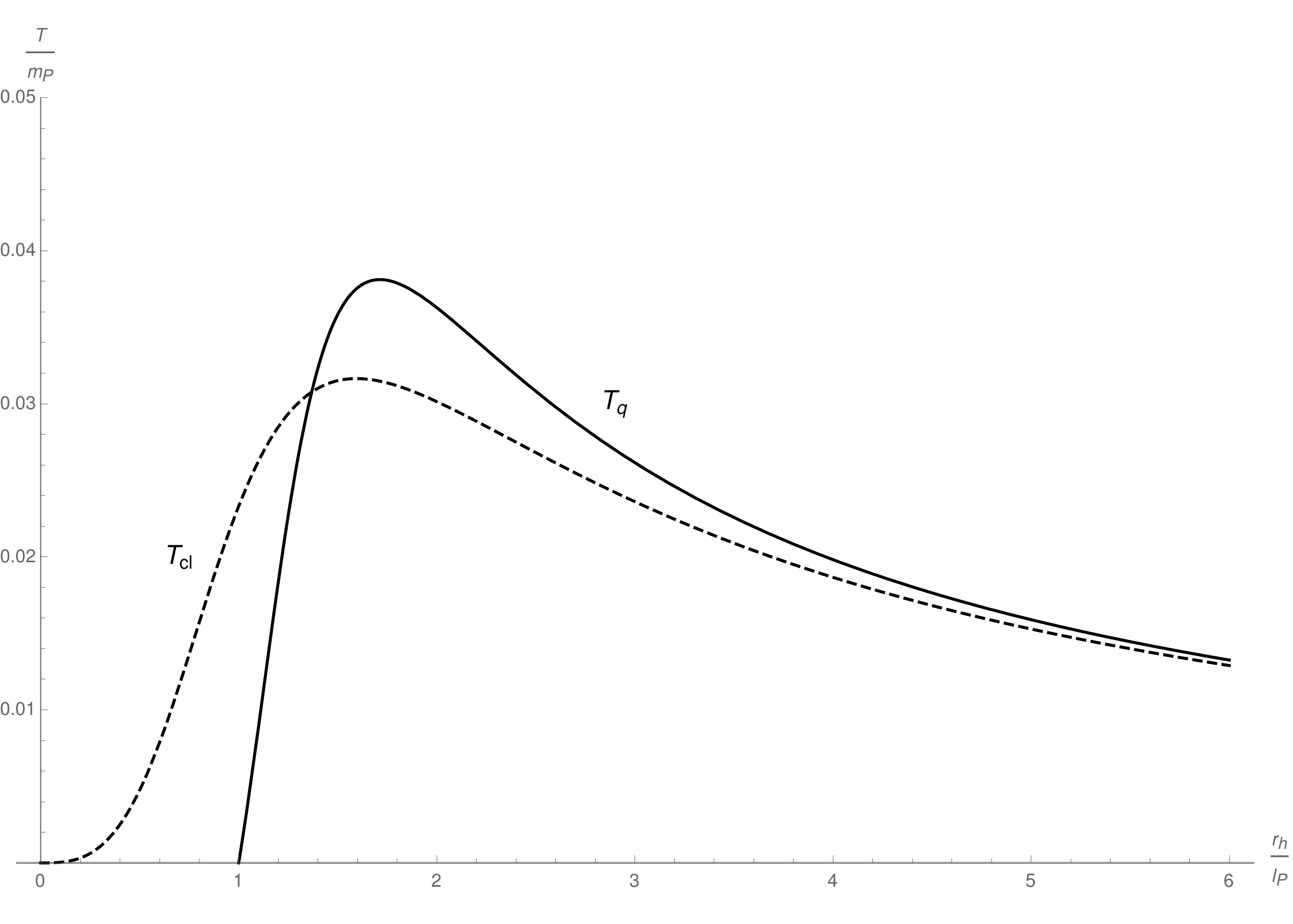}
\caption{Plot of the classical  temperature (\ref{tcl}), dashed line, and the one-loop temperature (\ref{th2}), continuous line.}
\label{cltemp}
\end{figure}
\end{center}

Equation (\ref{effrh1}) shows that the minimal value of $r_h$ is no more zero, but equal to $l_{Pl}$:

\begin{equation}
 \left(\, r_h\,\right)_{min}=l_{Pl}
\end{equation}

Therefore, any black hole has to be larger than $l_{Pl}$.\\
The corresponding metric function is given by 

\begin{equation}
 f(r)= 1 - \frac{\sqrt{2}m}{\mu_{Pl}}\left(\, 1 + \sqrt{1 + \left(\, \frac{r^2}{l_{Pl}^2}-1\,\right)^2}\,\right)^{-1/2}
\end{equation}

Once again, we recall that this semi-classical description has the same physical limitation, $r > l_{Pl}$, as in the 
tree-level case. Nevertheless, one finds a significant difference is considering the Hawking temperature in the latter
case Fig.(\ref{cltemp}).  

\begin{equation}
 T_H = \frac{1}{4\pi l_{Pl}^4}\frac{r_h\left(\, r_h^2 -l_{Pl}^2\,\right)}{ 1 
+ \sqrt{1 +  \left[\,\left(\,r_h/l_{Pl}\,\right)^2-1\,\right]^2 } }
\frac{1}{\sqrt{1 +  \left[\,\left(\,r_h/l_{Pl}\,\right)^2-1\,\right]^2   }}\label{th2}
\end{equation}

The expression (\ref{th2}) vanishes for $r_h\to l_{Pl}$, instead of $r_h\to 0$ as in the tree-level case. This 
is the main effect of the one-loop quantum corrections. This behavior confirms that near Planck scale the distinction
between particles and black holes fades away and the thermodynamical description is no more an adequate one.


\section{Summary and discussion }
\label{end}
In this paper we have described a possible formation of microscopic black holes through an Higgs-like mechanism operating at the 
Planck scale. The model we discussed contains only one scalar field. It is clear that further extensions, involving scalar
multiplets, gauge fields, etc. can be considered. However, our intention was to test the feasibility of this alternative approach
in the simplest possible framework, before attempting to account for phenomenological implications. From this perspective,
the choice of the Planck energy as the lower bound for black hole production is very traditional. Nothing prevents 
that in more elaborate models one can lower, or raise , this  threshold energy. \\
The preliminary results obtained in this simple model are encouraging to proceed towards more involved realizations of this
type of Higgs mechanism.\\
We have also shown that, near the threshold energy, one-loop effects play an important role and modify the behavior of the
Hawking temperature which vanishes as $r_h\to l_{Pl}$. It is important to notice that the very concept of temperature has
to be taken with caution as in this critical region there is no clear  distinction between particles and black holes.
This hybrid object is what we named ``~black particle~'', a quantum lump of energy with  a Compton wavelength and a gravitational 
radius which are of the same order of magnitude.\\
Keeping in mind all the  limitations  of the geometrical description, we found that the horizon entropy (\ref{lns}), 
even in the classical limit, ``recalls'' quantum effects in the form of a logarithmic correction to the Area Law.\\

\end{document}